\documentclass[prc,aps,floats,onecolumn,floatfix,showpacs,nofootinbib,%
superscriptaddress]{revtex4}

\usepackage{amsmath}
\usepackage{amssymb}
\usepackage{amsbsy}
\usepackage{amsfonts}
\usepackage{graphicx}
\usepackage{tabularx}
\usepackage{multirow}
\usepackage{lscape}
\usepackage{rotating}
\usepackage{pstricks}
\usepackage{feynmf}
\usepackage{tikz}
\usetikzlibrary{shapes.geometric, arrows}
\tikzstyle{startstop} = [rectangle, rounded corners, minimum width=3cm, minimum height=1cm,text centered, draw=black, fill=red!30]
\tikzstyle{io} = [trapezium, trapezium left angle=70, trapezium right angle=110, minimum width=3cm, minimum height=1cm, text centered, draw=black, fill=blue!30]
\tikzstyle{process} = [rectangle, minimum width=3cm, minimum height=1cm, text centered, draw=black, fill=orange!30]
\tikzstyle{decision} = [diamond, minimum width=3cm, minimum height=1cm, text centered, draw=black, fill=green!30]
\tikzstyle{arrow} = [thick,->,>=stealth]

\begin{document}

\title{Nuclear surface energy solving Hartree-Fock equations with Gogny interactions using Lagrange mesh}


\author{D. Davesne}
\email{davesne@ipnl.in2p3.fr}
\affiliation{Universit\'e Lyon 1,
             43 Bd. du 11 Novembre 1918, F-69622 Villeurbanne cedex, France\\
             CNRS-IN2P3, UMR 5822, Institut de Physique des 2 Infinis de Lyon}

\author{A. Pastore}
\email{alessandro.pastore@cea.fr}
\affiliation{ CEA, DES, IRESNE, DER, SPRC, F-13108 Saint Paul Lez Durance, France}

\author{J. Navarro}
\email{navarro@ific.uv.es}
\affiliation{IFIC (CSIC-Universidad de Valencia), Parque Cient\'{\i}fico, Catedr\'atico Jos\'e Beltr\'an 2, E-46.980-Paterna, Spain}


\begin{abstract}
Hartree–Fock equations for finite-range interactions in a slab of nuclear matter are presented and solved using an algorithm based on the Lagrange mesh method. This approach is faster and more efficient than the Numerov algorithm commonly used in the literature.
Thanks to the improved numerical accuracy, we were able to perform calculations with sufficiently large boxes to minimize the impact of Friedel oscillations on the final results, achieving a precision on the surface energy within a few dozens of keV.
Results are presented for several Gogny interactions that have not been previously discussed. In addition, the inclusion of the spin–orbit term is examined, showing a net reduction of 1.2-1.9 MeV in the surface energy.
\end{abstract}


\pacs{
    21.60.Jz 	
    21.65.-f 	
    21.65.Mn 	
}
 
\date{\today}


\maketitle

\section{Introduction}

According to the seminal work of Bohr and Wheeler~\cite{bor39} the mechanism of nuclear fission can be explained in terms of a simple liquid drop (LD) model, which is based on a balance between Coulomb repulsion and surface tension~\cite{swi51}. If the surface energy coefficient is too high, the fission barrier is largely overestimate fission barriers and fission will not occur at all~\cite{bjo80}.

Over decades, more and more sophisticated many-body methods based on effective nucleon-nucleon (NN) interactions have been developed to describe nuclear fission phenomena~\cite{sch16}, but the simple idea suggested in Ref.~\cite{bor39} remains valid.
In the case of an effective NN interaction, the surface energy coefficient $a_s$ is no longer a free parameter of the model, as in the LD model or more advanced micro-macro models (see Ref.~\cite{kra12}). However, one can still obtain a simple estimate using Hartree–Fock (HF) calculations \cite{Book:Ring1980}  in semi-infinite nuclear matter (SINM) \cite{cot78}.
In Ref.~\cite{rys19}, the authors performed a detailed analysis of Skyrme forces \cite{sky58}, showing a direct correlation between the fission barrier heights of a few selected fissile nuclei and the surface energy coefficient $a_s$.
For values of $a_s$ larger than 20 MeV, the fission barriers are too high and fission is strongly suppressed. For other effective NN interactions such as finite-range Gogny interactions \cite{dec80}, there are no so detailed studies on this topic, but the authors of Ref.~\cite{ber84} have discussed the failure of the Gogny D1 \cite{dec80} interaction in describing the fission barriers at very large deformations and they ascribed it to the too large value of $a_s$.
The calculation presented in Ref.~\cite{cot78} of this surface energy coefficient in SINM (confirmed in Ref.~\cite{dav23}) for D1 lead to $a_s=20.1$ MeV. Although a more systematic investigation should be done, we can argue that, following the analysis done Ref.~\cite{rys19} for Skyrme functionals, a surface coefficient of $\approx20$ MeV overestimate fission barriers and thus the corresponding interaction is not suitable to study fission observables.

The SINM calculations for Skyrme forces are relatively simple and they have been extensively performed within the scientific literature, see Refs.~\cite{far81,cen98,dan09,jod16,rys19,pro22} for example. However, once we deal with finite range forces, the exchange term becomes non-local and the resulting HF equations become integro-differential, thus making the computational task more demanding. The authors of Ref.~\cite{cot78,dav23} have presented a numerical procedure based on Numerov method~\cite{nou24} together with a localisation of the Fock potential (see Ref.~\cite{gra02,mic09} for details on this method).
This procedure is numerically time consuming since it requires a double convergence procedure: one for the fields and another one for the single-particle wavefunctions. In addition, in Ref.~\cite{dav25}, we have analysed the mean field potentials obtained with a variety of Gogny interactions and we have found that although the total potential at low momenta is very similar for all interactions, the relative balance between the direct and Fock terms can be very different.
For the D1 interaction for example, the Hartree plus density dependent term is close to zero and the full attraction arises only from the Fock potential, while for D3G3~\cite{bat22} the final potential arise from a strong cancellation between a strongly attractive Fock potential and strongly repulsive Hartree plus density dependent. For interactions with such a behaviour, the procedure presented in Ref.~\cite{cot78,dav23} breaks down so that a new numerical approach is actually required.

The focus of the present article is thus to present a new and more efficient  numerical procedure to solve the HF equations in SINM for a variety of Gogny interactions, namely the Lagrange mesh method~\cite{bay15}. This is the first step in order to perform a study as rigorous as the one of Ref.~\cite{rys19}. The article is organised as follows: in Sect.~II we recall the INM properties of the selected Gogny interactions. In Sect.~III we present
the formalism of the Hartree–Fock equations for a  SINM slab, then we describe the Lagrange mesh algorithm and we detail the resulting equations. In Sect.~IV we present our results and the conclusions are drawn in Sect.~V.

\section{Gogny interactions}\label{sec:iNM}

The Gogny interaction  in its \emph{standard} form is composed of a central $V_C$, density dependent $V_{DD}$ and a spin orbit $V_{SO}$ term as

\begin{equation}\label{eq:Gogny:standard}
V(\mathbf{r}_1,\mathbf{r}_2)=V_C(\mathbf{r}_1,\mathbf{r}_2)+V_{DD}(\mathbf{r}_1,\mathbf{r}_2)+V_{SO}(\mathbf{r}_1,\mathbf{r}_2).
\end{equation}

\noindent The various terms read
\begin{eqnarray}
V_C(\mathbf{r}_1,\mathbf{r}_2)&=&\sum_i (W_i+B_iP_\sigma-H_iP_\tau-M_iP_\sigma P_\tau)e^{-r_{12}^2/\mu_i^2}\,,\\
V_{DD}(\mathbf{r}_1,\mathbf{r}_2)&=&t_3(1+x_3P_\sigma) \rho^\gamma \delta(\mathbf{r}_{12})\,,\label{eq:DD}\\
V_{SO}(\mathbf{r}_1,\mathbf{r}_2)&=&i W_0 (\mathbf{k}' \times \mathbf{k})(\sigma_1+\sigma_2) \delta(\mathbf{r}_{12})\,.
\end{eqnarray}
$P_\sigma,P_\tau$ are the spin/isospin exchange operators,  $\mathbf{r}_{12}=\mathbf{r}_1-\mathbf{r}_2$, $\mathbf{k}=-\frac{i}{2}(\nabla_1-\nabla_2)$ is the relative momentum operator acting on the right and $\mathbf{k}'$ its conjugate acting on the left, and $\sigma_i$ are the Pauli matrices. See Refs~\cite{dav16,dav21} for more details.
The density-dependent term, $V_{DD}(\mathbf{r}_1,\mathbf{r}_2)$, and the spin-orbit one, $V_{SO}(\mathbf{r}_1,\mathbf{r}_2)$, have the same form as in the Skyrme interaction both being zero-range~\cite{sky58}.
A second density-dependent term of the form $V_{DDb}(\mathbf{r}_1,\mathbf{r}_2)=t_{3b}(1+x_{3b}P_\sigma) \rho^{\gamma_b} \delta(\mathbf{r}_{12})$ has been considered for the D1P parametrisation~\cite{far99}. 
In recent years, we can find in the scientific literature new forms of the Gogny interaction having also a finite-range density dependent term~\cite{cha15} or a finite-range tensor~\cite{cha15,ots06}. Since the inclusion of these extra terms requires major modifications of the HF equations, we will concentrate in this paper only on the Gogny interactions with the above \emph{standard} form.

The pool of Gogny interactions discussed in the current article is presented in Tab.~\ref{tab:eos} and we refer the interested reader to  Refs.~\cite{sel14,dav25} for a brief description of each interaction as well as the results they produced when used to perform HF calculations in infinite nuclear matter (INM).
In Tab.~\ref{tab:eos} we have reported some relevant INM properties such as the energy per particle at saturation density, the saturation density, the isoscalar effective mass $m^*/m$, the symmetry energy $J$ and its first derivative $L$.
We refer to Ref.~\cite{sel14} for a detailed definition of these quantities.

\begin{table}[h!]
    \centering
    \begin{tabular}{l|c|c|c|c|c|c}
    \hline
        \hline 
         &  E/A($\rho_0$) [MeV] & $\rho_0$ [$\text{fm}^{-3}$ ]& $m^*/m$ & $J$ [MeV] & $L$ [MeV] \\
        \hline
        D1~\cite{dec80}  & -16.30&0.166 & 0.670 & 30.7 & 18.3 \\
        D1S~\cite{dec80} & -16.01 &0.163 & 0.697 & 31.1 & 22.4\\
        D1N~\cite{chap08} &-15.96 & 0.161  & 0.747 & 29.6 & 33.6\\
        D1M~\cite{gor09d1m} & -16.02& 0.165  & 0.746 & 28.5 & 24.8\\
        D1M*~\cite{gon18} &-16.06 & 0.164  & 0.746 & 30.2 & 43.1\\
        D3G3~\cite{bat22} &-16.05 & 0.164  & 0.678 &32.5  & 36.7\\
        D3G3M~\cite{bat24}  & -16.06 & 0.164 & 0.739 & 28.5 & 25.4 \\
        D1P ~\cite{far99} &-15.04 & 0.169  & 0.672 &32.4  & 49.7\\
        D250 ~\cite{bla95} &-15.84 & 0.158 & 0.702 & 31.6 &24.8 \\
        D260 ~\cite{bla95} &-16.25 &0.160  & 0.615 & 30.1 &17.6 \\
        D280 ~\cite{bla95} &-16.33 & 0.153 & 0.575 & 33.1 & 46.5\\
        D300 ~\cite{bla95} &-16.22 & 0.156 & 0.681 & 31.2 & 25.8\\
         \hline
    \end{tabular}
    \caption{INM properties of selected Gogny interactions. See text for details.}
    \label{tab:eos}
\end{table}

We notice that since all these interactions have been adjusted to reproduce some target values in INM around saturation density, we do observe a very little variance in the obtained results, except for $L$. Just as a comparison we refer the reader to see the Skyrme case discussed for example in Ref.~\cite{dut12}.

\section{Semi infinite nuclear matter}\label{Sec:SINM}

Following the work of Swiatecki~\cite{swi51}, we define SINM as an infinite medium along two directions, say  $x,y$ in cartesian coordinates, and with a well defined surface along the $z$-axis. Along this direction, the matter density $\rho$ varies between two asymptotic values 
\begin{eqnarray}\label{eq:rho:sinm}
\lim_{z\rightarrow \pm \infty} \rho(z)=\left\{ \begin{array}{c}
\rho_0\,,\\
0\,.\end{array} \right.
\end{eqnarray}
Since the system is infinite along $z$, the  value $\rho_0$ is not impacted by the presence of a surface around $z=0$ and is thus set to the standard INM value and reported in Tab.~\ref{tab:eos} for each Gogny interaction.
Following the idea of Ref.~\cite{bon76}, we replace Eq.~(\ref{eq:rho:sinm}) with the profile of a slab within an infinite box of size $[-L,L]$ along the $z$ direction.

Since we are going to solve the HF equations in a self consistent procedure, we use as a starting potential $U_{\rm start}(z)$ the following expression
\begin{eqnarray}
    U_{\rm start}(z)=\left\{ \begin{array}{cc}
\frac{\displaystyle U_0}{\displaystyle 1+{\rm e}^{(z-L_0)/d}} & z\ge 0\,,\\
\frac{\displaystyle U_0}{\displaystyle 1+{\rm e}^{-(z+L_0)/d}}&z<0.\end{array} \right.
\end{eqnarray}
with $U_0=-52$ MeV, $L_0=3 L/4$ and $d=0.6$ fm. These parameters allow to place the matter within the slab and to obtain a value of the density reasonably close to the final solution. Although strictly speaking these parameters do not impact the final result, it is anyhow important to have a reasonable starting point in order to avoid numerical issues.

The total energy of  the system reads
\begin{equation}
    E_L=\int_{-L}^L \mathcal{E}(z)dz,
\end{equation}
where $\mathcal{E}(z)$ is the energy density~\cite{cot78}. We then calculate the same quantity, but considering that we have INM in the whole range. In that case, we obtain
\begin{equation}
    E'_L=a_v\int_{-L}^L \rho(z)dz,
\end{equation}
where $a_v={E/A}$ is the energy per particle of INM at saturation density $\rho_0$. By definition, the difference between these two quantities represents the surface energy per nucleon and leads immediately to
\begin{eqnarray}
  a_s=\left(\frac{36 \pi}{\rho_0^2}\right)^{\frac{1}{3}} \lim_{L\rightarrow \infty}\int_{-L}^L  \left(\mathcal{E}(z)-a_v\rho(z)\right)dz\,.
\end{eqnarray}

The goal is now to determine $\mathcal{E}(z)$. Therefore we now discuss the Hartree-Fock equations. Since in Ref.~\cite{dav23}, we have already provided a detailed presentation, we just sketch here the relevant equations and highlight the modifications we introduced due to the new numerical method.
 Since the system is infinite and homogeneous in $x,y$ plane, it is convenient to use a mixed representation of the wavefunction in coordinate and momentum spaces~\cite{sto70} as
\begin{eqnarray}\label{wave}
\phi_\mu(\mathbf{r})= \sqrt{2} \psi_{\lambda}(z,k_z,k_t) \chi_\lambda(\hat{\mathbf{k}}_t)e^{i \mathbf{k}_t\cdot \mathbf{r}}\,,
\end{eqnarray}
where $\mathbf{k}_t$ is the transverse momentum (with respect to the surface).  The spinor $ \chi_\lambda(\hat{\mathbf{k}}_t)$ is an eigenvector of $(\mathbf{k} \times \boldsymbol{\sigma})_z$ with eigenstates $\lambda=\pm1$.
In this representation, the Hartree-Fock equations read 
\begin{equation}\label{eq:HF}
\left( -\frac{\hbar^2}{2m}\frac{d^2}{dz^2}+U^0_q(z)+\lambda k_t U_q^{SO}(z)-\varepsilon_{q}(k^{eq}_z,k_t)+\frac{\hbar^2 k_t^2}{2m}\right)\psi_{q\lambda}(z,k^{eq}_z,k_t)=\int_{-\infty}^{\infty}U^1_q(z,z',k_t)\psi_{q\lambda}(z',k^{eq}_z,k_t)dz'\;,
\end{equation}
where $q=n,p$ is the isospin index. 
$U^0_q,U_q^{SO},U_q^1$ represent the single particle potentials, whose explicit expressions are given later on.
Notice that as compared to Ref.~\cite{dav23}, we have replaced $k_z$ in Eq.~(\ref{eq:HF}) with an \emph{equivalent} $k_z^{eq}$. The reason will become clear in the next section but it is worth noticing at this stage that, contrary to the case discussed by Cot\'e and Pearson \cite{cot78}, the eigenvalues $\varepsilon_{q}(k^{eq}_z,k_t)$ do not form anymore a continuum of states corresponding to the values of the infinite medium  $\varepsilon_q^{INM}$, but they are discretised by the choice of placing the slab in a box with vanishing boundary conditions.
As a consequence we do  not know \emph{a priori} the value of $k_z^{eq}$ before actually solving Eq.~(\ref{eq:HF}).
Once the eigenvalues $\varepsilon_{q}(k^{eq}_z,k_t)$ are determined,  $k_z^{eq}$ is then obtained as the solution of the following non-linear equation
\begin{eqnarray}\label{eq:kzequiv}
    \varepsilon_q^{INM}=\frac{\hbar^2}{2m}\left[(k^{eq}_z)^2+k_t^2\right]+U^{INM}_q\left(\sqrt{(k^{eq}_z)^2+k_t^2}\right)
\end{eqnarray}
where $U^{INM}_q$ is the single particle potential in INM. See Ref.~\cite{dav25} for the explicit expressions of these quantities.
Within the Hartree–Fock approximation, all single-particle states are filled up to the Fermi momentum $k_{Fq}=(3\pi^2\rho_q)^{1/3}$ for each species. Therefore, Eq.~(\ref{eq:kzequiv}) together with the relation $\sqrt{(k^{eq}_z)^2+k_t^2}\le k_F^q$ allow us to select the eigenvalues used in the following to build the HF densities and the fields.%

\subsection{Lagrange mesh}

\noindent We summarize here the Lagrange mesh method~\cite{bay15} employed to solve the HF integro-differential equation~\eqref{eq:HF}. 
We consider a set of real functions $f_i(x)$ that are infinitely differentiable and a mesh  of N points $z_i = x_i \; dz $ over an interval $[a,b]$ together with weights $\lambda_i$ of a given Gauss quadrature. These functions have to satisfy the following conditions
\begin{eqnarray}
    f_i(x_j)&=&\lambda_i^{-1/2}\delta_{ij}\\
    \langle f_i| f_j\rangle&=& \int_a^b f_i(x)f_j(x)dx=\langle f_i| f_j\rangle_G \, ,
\end{eqnarray}
where the last equality states that the Gauss quadrature is exact for the product of Lagrange functions.
The two previous conditions have as a consequence that the Lagrange functions are orthonormal since
\begin{equation}
    \langle f_i| f_j\rangle_G=\sum_{k=1}^N \lambda_k  \lambda_i^{-1/2}\delta_{ik}\lambda_j^{-1/2}\delta_{jk}=\delta_{ij}
\end{equation}

In order to get the solution of Eq.~(\ref{eq:HF}), one has to expand (for each set of $k^{eq}_z,k_t$) the wavefunctions $\psi_{q\lambda}$ in Lagrange functions as
\begin{equation}
    \psi_{q\lambda}(z,k^{eq}_z,k_t)=\frac{1}{\sqrt{2dz}}\sum_{j=1}^N c_j(k^{eq}_z,k_t)f_j(z)
\end{equation}
With a uniform mesh in the $[-L,L]$ range defined as
\begin{equation}
    z_i =-\frac{1}{2}(N-1)dz,\dots,\frac{1}{2}(N-1)dz
\end{equation}
with $dz=2L/N$, the HF equations read
\begin{equation}\label{eq:HF:lagrange}
    \sum_{j=1}^N\left[\frac{\hbar^2}{2m}T_{ij}+\left(U_q^0(z_i)+\lambda k_t U_q^{SO}(z_i)+\frac{\hbar^2 k_t^2}{2m} \right)\delta_{ij}-(\lambda_i \lambda_j)^{1/2} U^1_q(z_i,z_j)dz \right]c_j=\varepsilon(k_z^{eq},k_t)c_j,
\end{equation}
where $T_{ij}$ is the matrix element of the kinetic energy. A very striking result of this derivation is that the integral over the Fock potential in Eq.~(\ref{eq:HF}) disappears since we need to evaluate the non-local potential at the point of the grid without passing for an integration (see Ref.~\cite{bay15}). This is a remarkable simplification of the problem since Eq.~(\ref{eq:HF:lagrange}) can be solved via a simple diagonalisation without requiring an iterative procedure to identify the eigenvector as presented in Refs.~\cite{cot78,dav23}.

In order to specify the kinetic energy matrix element, we have to chose an appropriate set of Lagrange functions $f$. According to Ref.~\cite{bay15}, the most adapted to deal with the current problem is the Lagrange-Fourier set
\begin{equation}
    f_j(z_i)=\frac{\sin \left[ \pi (i-j) \right] }{N\sin \left[ \frac{\pi}{N}(i-j) \right] }.
\end{equation}

The quadrature weight of these functions is $\lambda_i=1$ and the kinetic  matrix element reads
\begin{eqnarray}
T_{ij}=\left\{ \begin{array}{cc}
(-1)^{i-j} \frac{2\pi^2}{N^2dz^2} \frac{\cos[\pi (i-j)/N]}{\sin^2[\pi (i-j)/N]} & i\ne j \\
\frac{\pi^2}{3 dz^2}\left( 1-\frac{1}{N^2}\right) & i=j
\end{array}\right.
\end{eqnarray}

For this specific choice of basis functions, the matrix elements are computed exactly. As a result, we are no longer constrained to use very small step sizes to preserve the accuracy of the derivatives, as was necessary in our previous work \cite{dav23}. Using the solutions of Eq.~(\ref{eq:HF:lagrange}), we can write the matter, kinetic and spin-current densities as
\begin{eqnarray}
    \rho_q(z)&=&\frac{1}{\pi} \sum_\lambda \int_0^{k_{Fq} } \sum_{k_z^{eq}} |\psi_{q\lambda}(z,k^{eq}_z,k_t)|^2 k_t dk_t  \label{eq:rho}\\
    \tau_q(z)&=& \frac{1}{\pi} \sum_\lambda \int_0^{k_{Fq} }  \sum_{k_z^{eq}} \left[|\psi_{q\lambda}(z,k^{eq}_z,k_t)|^2 k^2_t +|\psi^\prime_{q\lambda}(z,k^{eq}_z,k_t)|^2\right]dk_t  \label{eq:tau}\\
    J(z)&=&\frac{1}{\pi} \sum_\lambda \lambda \int_0^{k_{Fq} } \sum_{k_z^{eq}} |\psi_{q\lambda}(z,k^{eq}_z,k_t)|^2 k^2_t dk_t 
\end{eqnarray}

The integral over $k_t$ has been discretised using a Gauss-Legendre (GL) quadrature. In the following section we discuss the dependence of the numerical accuracy on the adopted number of points $N_{GL}$.
The first derivative of the wave function appearing in Eq.~(\ref{eq:tau}) is determined with the matrix $D_{ij}$ below, applied to the set of our wavefunctions 
\begin{eqnarray}
D_{ij}=\left\{ \begin{array}{cc}
(-1)^{i-j} \frac{\pi}{N\sin [(i-j)/N]} & i\ne j \\
0 &i=j
\end{array}\right.
\end{eqnarray}
Finally, as already anticipated, we have replaced the integral over $k_z$ by a sum over the allowed values of $k_z^{eq}$ as explained before. The same applies to the Fock field that now reads
\begin{eqnarray}
U^1_q(z,z',k_t)&=&\frac{1}{2}\sum_i \sum_{q'}\sum_{\lambda} \mu_i^2 e^{-(z-z')^2/\mu_i^2} \left[(2M_i+H_i)-(2B_i+W_i)\delta_{qq'} \right] \nonumber \\
&\times& \int_0^{k_{Fq}}  \sum_{k_z^{eq}} k'_t \psi_{q'\lambda}^{*}(z,k_z^{eq},k_t')\psi_{q'\lambda}(z',k_z^{eq},k_t')e^{-\mu_i^2(k_t^2+k_t^{'2})/4} I_0\left( \frac{\mu_i^2 k_tk_t'}{2}\right)dk_t' ,\label{eq:fieldsF}
\end{eqnarray}
where $I_0(x)$ is the modified Bessel function of zeroth order \cite{abr88}.
This new expression allows for a significant reduction of computational cost since the sum is now exact.
The other fields do not change, but we give their expression here for completeness\footnote{In Ref.~\cite{dav23} there is a mistake in the Eq.~(13) relative to the density dependent terms. }
\begin{eqnarray}
U^0_q(z)&=&\frac{\pi}{2}\sum_i \mu_i^2 \int_{-\infty}^{\infty} \left[ (2W_i+B_i)\rho(z')-(2H_i+M_i)\rho_q(z')\right]e^{-(z-z')^2/\mu_i^2} dz'-\frac{W_0}{2}\frac{d}{dz}\left( J(z)+J_q(z) \right)\nonumber \\
&+& t_3 \rho^\alpha(z)\left[\left( 1+\frac{x_3}{2}\right)\rho(z)- \left( \frac{1}{2}+x_3\right)\rho_q(z)\right]+\frac{t_3}{2}\alpha \rho^{\alpha-1}\left[ \left(1+\frac{x_3}{2} \right)\rho^{2}-\left(x_3+\frac{1}{2} \right) \sum_q \rho_q^2 \right] \\
U_q^{SO}(z)&=&\frac{1}{2}W_0\left( \nabla \rho(z)+\nabla\rho_q(z)\right). \label{eq:fieldsSO}
\end{eqnarray}

The fields are calculated self-consistently and used to solve Eq.~(\ref{eq:HF:lagrange}) until convergence is achieved. 
We can then calculate the total energy density of the system as
\begin{eqnarray}\label{eq:energydensity}
\mathcal{E}(z) =\frac{1}{2} \sum_q \left\{\varepsilon_q(z) +\frac{\hbar^2}{2m}\tau_q(z)\right\}-U_{\rm{rearr}}(z)\rho(z)
\end{eqnarray}
where
\begin{eqnarray}
\varepsilon_q(z)&=& \frac{1}{\pi} \sum_\lambda \int_0^{k_{Fq} } \sum_{k_z^{eq}}  \varepsilon_{q\lambda} (k_z,k_t)|\psi_{q\lambda}(z,k^{eq}_z,k_t)|^2 k_t dk_t \\
U_{\rm rearr}(z)&=&\alpha t_3 \rho(z)^{\alpha-1}\left[ \frac{1}{2}\left(1+\frac{x_3}{2} \right)\rho(z)^2-\frac{1}{4} \left(1+2 x_3 \right)\sum_q\rho_q(z)^2 \right]
\end{eqnarray}
$U_{\rm rearr}(z)$ is a rearrangement term arising from the explicit density dependence of the Gogny interaction.\footnote{Notice there is a typo in Eq.~(3.26) of Ref.~\cite{cot78}.}

In the previous derivation, we have explicitly neglected the effects arising from the presence of a residual pairing interaction. Previous studies \cite{baldo2000microscopic,baldo1999surface} have shown that the pairing field is peaked at the surface of SINM, we thus may expect that it could have an impact on the diffusivity of the density and thus on the values of the surface energy. An accurate study would require to describe the system using Hartree-Fock-Bogoliubov (HFB) equations \cite{Book:Ring1980}. To date, this remains computationally too expensive and we postpone the discussion of the effect of pairing field on $a_s$ in a future work.

\section{Results}\label{sec:results}

\begin{figure}[h!]
    \centering
        \centering
          \includegraphics[width=0.6\textwidth]{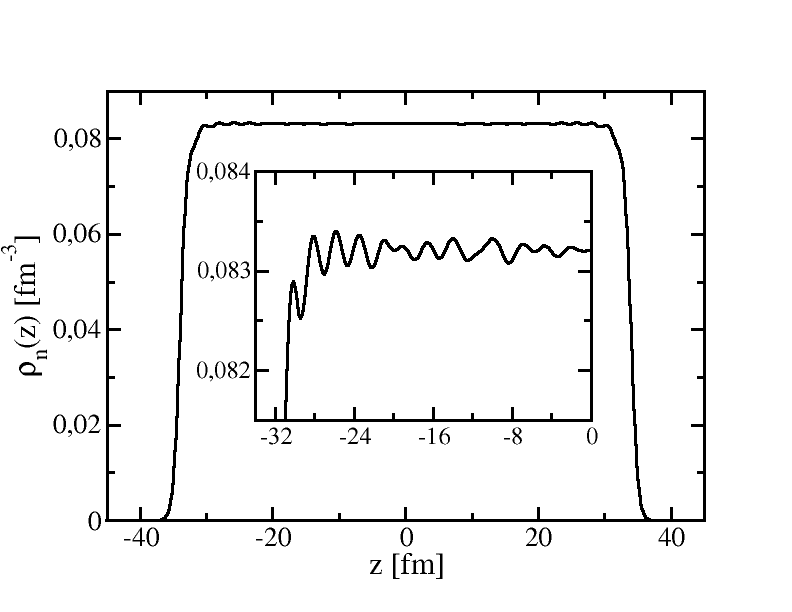}
        \caption{Neutron density obtained in a full HF calculation using D1 interaction (without spin-orbit term). The inset shows the presence of Friedel oscillations.}
    \label{fig:rhoN}
\end{figure}

The presence of a surface induces the appearance  of Friedel oscillations whose typical length scale $L_F$ is of order $\pi/k_F \simeq 2.3$ fm at saturation density.
As an example in Fig.~\ref{fig:rhoN}, we show the neutron density obtained in a HF calculation of SINM using the Gogny D1 interaction (without spin-orbit term). We notice that despite a very large box ($L=46$ fm) the Friedel oscillations extend from the surface up to the center of the slab.
Furthermore, compared to the results presented in Ref.~\cite{jod16}, we also notice an interference pattern in the Friedel oscillations, related to the fact that we work with a finite slab and not a piece of SINM. Ideally, the center of the slab should exhibit the asymptotic behaviour of infinite matter and thus should be free of these oscillations. However, depending of the box size, the center actually corresponds to a maximum or a minimum of an oscillation. In order to further reduce the sensitivity of this effect, we average the value of $a_s$ over 4 fm, which is an interval larger that the typical length of the Friedel oscillation ($\approx \pi/k_F$), starting from the centre of the slab : using such a procedure we observe a better convergence of $a_{s}$.

\begin{figure}
    \centering
        \centering
          \includegraphics[width=0.45\textwidth]{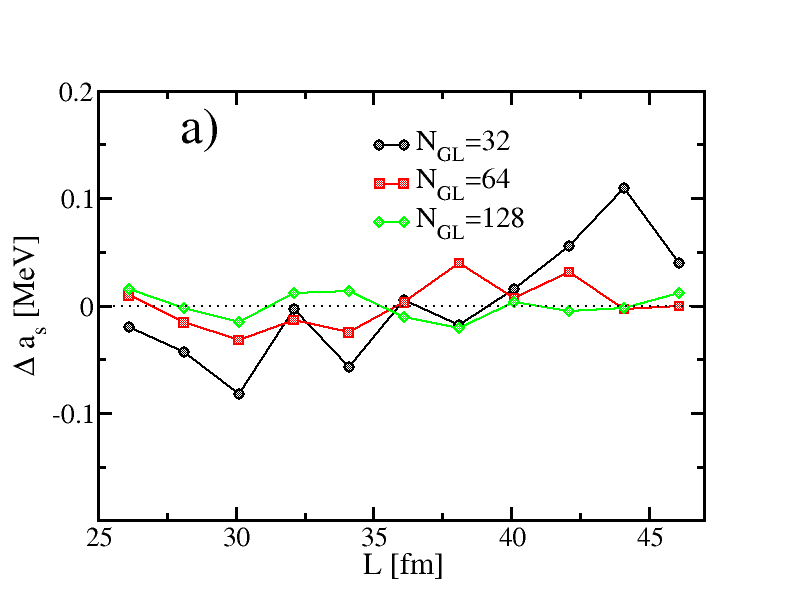}
        \includegraphics[width=0.45\textwidth]{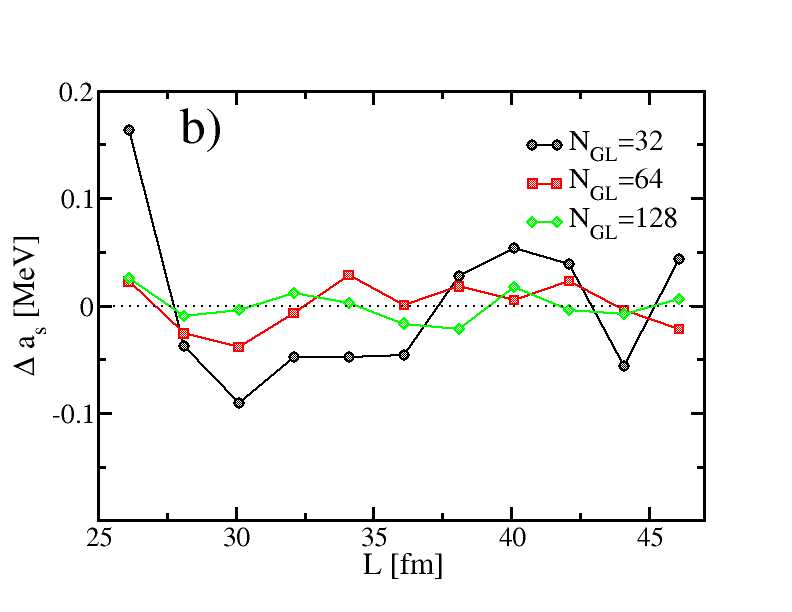}
        \caption{Evolution of the surface energy coefficient $a_s$ for the Gogny D1 interaction without (panel a) and with spin-orbit term (panel b) as a function of half of the size of the slab ($L$) and for various number of integration points in the $k_t$ direction. The dotted lines represent the zero and are there just to guide the eye.  See text for details. }
    \label{fig:convergence}
\end{figure}

To extract a reliable value, we need furthermore to consider $L \gg L_F$ and prove that the results are not too sensitive to the value of $L$. In Fig.~\ref{fig:convergence}, we show the  dependence of the discrepancy $\Delta a_s=a_s-\bar{a}_s$, where $\bar{a}_s$ is obtained by averaging the results of different calculations with different values of $L$ ranging from 26 fm to 46 fm.
The value of $\Delta a_s$ is calculated as a function of the box and for various choices of $N_{GL}$ along the $k_t$ direction. 
We notice 
that in order to have a precise result, \emph{i.e.} with an accuracy less than 50 keV one needs a value of at least $N_{GL}=64$ and to use boxes larger than 30 fm.
For all these calculations, we fixed the mesh size at $dz=0.2$ fm. We tested that reducing it to $dz=0.1$ fm the results do not change.

In Tab.~\ref{tab:asurf}, we report the values of $a_s$ obtained with and without spin-orbit term. As mentioned before, in order to further reduce the dependence on $L$, we decided to calculate the average value over the various box sizes. 
In doing so, we can calculate also the root mean square deviation for our results and thus provide an estimate of the error bar of these calculations, which in all cases is roughly at the $0.2\%$ level.

\begin{table}[!h]
    \centering
    \begin{tabular}{c|c|c}
    \hline
        \hline 
         & $\bar{a}_s$[MeV]  &$\bar{a}_s$[MeV] (no s.o.) \\
        \hline
        \hline
              D1 &19.74 $\pm$ 0.02 & 20.92 $\pm$ 0.02\\
        D1S & 17.91 $\pm$ 0.01 &19.35 $\pm$ 0.02\\
        D1N &18.04 $\pm$ 0.01  & 19.29 $\pm$ 0.02\\
        D1M &18.14 $\pm$ 0.01 & 19.44 $\pm$ 0.01\\
        D1M* &18.20 $\pm$ 0.01 & 19.60 $\pm$ 0.01\\
        D3G3 &17.83 $\pm$ 0.02 & 19.08 $\pm$ 0.02\\
        D3G3M &18.37 $\pm$ 0.02 & 19.84 $\pm$ 0.02\\
        D1P &17.06 $\pm$ 0.02 & 18.97 $\pm$ 0.02\\
        D250  &18.61 $\pm$ 0.02 & 20.07 $\pm$ 0.02\\
        D260  &19.54 $\pm$ 0.02 & 20.95 $\pm$ 0.02\\
        D280  &19.77 $\pm$ 0.02 & 21.08 $\pm$ 0.02\\
        D300  &19.58 $\pm$ 0.02 & 21.18 $\pm$ 0.02\\
         \hline
         \hline
    \end{tabular}
    \caption{The average surface energy coefficient for various Gogny interactions. In the last column we provide the results removing the spin orbit term. See text for details.}
    \label{tab:asurf}
\end{table}

\begin{figure}
    \centering
        \centering
         \includegraphics[width=0.55\textwidth]{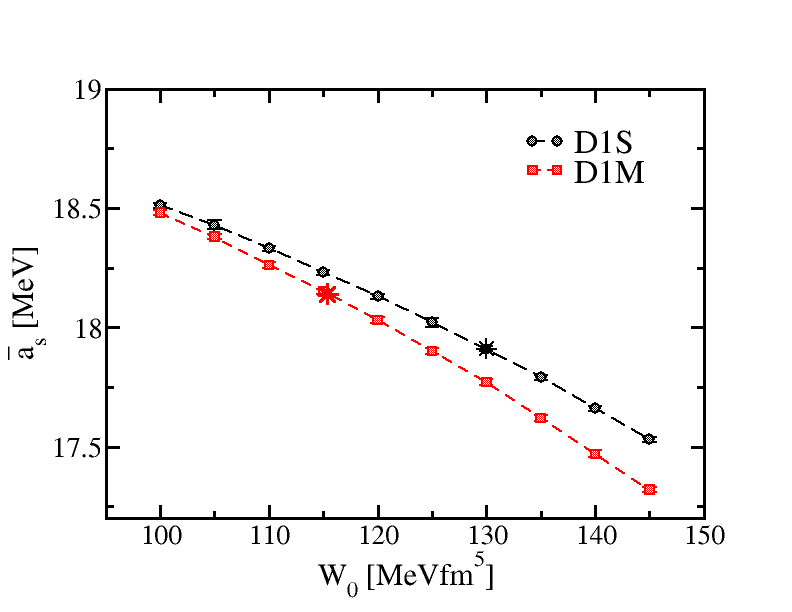}
        \caption{Evolution of the surface energy coefficient $\bar{a}_s$ in function of the spin-orbit parameter $W_0$ for two selected Gogny interactions. The star symbols indicate the nominal value of $W_0$ for each interaction. See text for details.}
    \label{fig:W0}
\end{figure}

We observe that the interactions D1, D260, D280 and D300 have a surface energy coefficient larger than 19 MeV so most likely, these interactions are not suitable to describe fission process. All other Gogny interactions have a value around 18 MeV which should allow for a reasonable description of fission barriers \cite{rys19}. Of course this conclusion is rather speculative since no systematic study  exists, although we know that interactions as D1S and D1M (D1M*) are routinely used to describe this physics \cite{rod14,rod20,gou05}. Thus, we  assume that the surface coefficient should not too be large to reduce (or forbid) fission.
Finally, the smallest surface energy coefficient is found for the D1P interaction. This interaction shows noteworthy features in infinite nuclear matter \cite{dav25}, possibly due to the presence of an additional density-dependent term that mitigates parameter correlations, even though it does not yield satisfactory results for finite nuclei.
In all cases we observe that the surface tension is mainly determined by the bulk properties of the interaction and the presence of the spin-orbit term contributes to a reduction of roughly 1.2 - 1.9 MeV.
As a simple analysis to verify the role of the spin-orbit on $\bar{a}_s$, we decided to perform a series of calculations for D1S and D1M interactions as a function of the spin-orbit parameters $W_0$ only.
Of course this is only a crude approximation since we know that the parameters of the interaction are correlated with each other (see Ref.~\cite{bec19} for a short discussion), but it is still interesting to see how strong the correlation is between these two quantities.
In Fig.~\ref{fig:W0}, we show the evolution of $\bar{a}_s$ for different values of $W_0$, leaving the other parameters unchanged. We recall that the values for the spin-orbit parameters  are $W_0=130$ MeVfm$^5$ and $W_0=115.36$ MeVfm$^5$, respectively for D1S and D1M (and marked with a star on the figure).
We observe a linear trend for small variations of $W_0$ around its nominal value although the slope of the curve depends on the adopted interaction. This feature may be exploited during fitting protocols in order to fine tune the results, although a rigorous study of the complete covariance matrix of the interaction would be still necessary.

\section{Conclusions}\label{sec:conclusion}

We have discussed the numerical advantages of using the Lagrange mesh to solve the Hartree-Fock equations in semi-infinite nuclear matter for a variety of Gogny interactions.
Thanks to method presented here it is now possible to have a complete set of results for Gogny interactions, including the most recent ones, that paves the way to a systematic study of fission barriers~\cite{rys19}. The main advantage of using the Lagrange mesh is that the Fock field has to be evaluated over the point of the grid and not integrated, thus improving remarkably the accuracy of the solution. Moreover the first and second derivatives are exact~\cite{bay15}.
This technique allows to further extend the size of the box used compared to previous calculations~\cite{cot78,dav23} and thus reduces the importance of Friedel oscillations on the final results.
We have been thus able to prove that our results do not depend on the size of the slab or on the number of integration points used to evaluate the integrals appearing in our equations.
We have also studied the role of the spin-orbit term and its contribution to the surface energy coefficient and we find that this is roughly 1.2-1.9 MeV for all interactions considered here.
By freezing all parameters of the interaction and varying the spin orbit term we have observed a strong linear correlation between the spin-orbit parameter $W_0$ and $\bar{a}_s$.

Finally, it is worth mentioning that new Gogny interactions including additional tensor terms have been recently developed~\cite{gra13,zie23} and given the strong correlation observed between spin-orbit and tensor term~\cite{les07}, we may expect an impact on the surface energy.
The inclusion of these additional terms is not straightforward since the simplification induced by the zero-range spin orbit term would not hold anymore thus implying a rewriting of the entire formalism.
We leave this analysis for a future work.

We conclude by stressing that the entire formalism presented here neglects the role of pairing effects on the surface energy. It is well known that pairing field in SINM is surface peaked~\cite{baldo2000microscopic} and it could also contribute to the determination of $a_s$, but a fully consistent treatment of pairing via a complete Hartree-Fock-Bogoliubov approach remains, although very interesting, very challenging.

\section{Acknowledgments}

The authors would like to thank M. Bender for 
bringing the Lagrange mesh method to their attention and his valuable support during the development of the numerical code. We also thank D.J Baye for kindly answering our questions on 
this method.

\bibliography{biblio}

\end{document}